\documentclass[11pt]{article}
\usepackage{latexsym}
\usepackage{amssymb}
\usepackage{epsf}
\usepackage{amsmath}
\usepackage{slashed}
\usepackage[hypertex]{hyperref}
 \usepackage[dvipdfm]{graphicx}

\newcommand{\ro}[1]{\sqrt{\mathstrut #1}}

\begin{document}
\pagestyle{empty}

\begin{flushright}
KEK-TH-1960
\end{flushright}

\vspace{3cm}

\begin{center}

{\bf\LARGE Gravitino thermal production revisited and a new cosmological scenario of gauge mediation}
\\

\vspace*{1.5cm}
{\large 
Hiraku Fukushima$^{1,2}$ and Ryuichiro Kitano$^{1,3}$
} \\
\vspace*{0.5cm}

$^1${\it KEK Theory Center, Tsukuba, Japan}\\
$^2${\it Department of Physics, Tohoku University, Sendai, Japan} \\
$^3${\it The Graduate University for Advanced Studies (Sokendai),
Tsukuba, Japan}
\vspace*{0.5cm}

\end{center}

\vspace*{1.0cm}

\begin{abstract}
{\normalsize We present a new scenario of gravitino dark matter which is
compatible with the thermal leptogenesis.  We confirm by an explicit
calculation in supergravity that the relic abundance of thermally
produced gravitino becomes insensitive to the reheating temperature once
the temperature of the Universe exceeds the mass scale of the messenger
fields.
In such a situation,
the correct baryon to dark matter ratio can be obtained by thermal leptogenesis
when the reheating temperature after inflation is high enough.
We demonstrate in a concrete model of gauge mediation that the correct
abundance of gravitino 
and baryon asymmetry can be reproduced
by considering the late-time entropy
production from the decay of the SUSY-breaking pseudo-moduli field. The
scenario is realized when the gravitino mass is $100~{\rm MeV} \lesssim
m_{3/2} \lesssim 1~{\rm GeV}$, and the messenger mass scale is
$10^6~{\rm GeV} \lesssim M_{\rm mess} \lesssim 10^9~{\rm GeV}$.  }
\end{abstract} 

\newpage
\baselineskip=18pt
\setcounter{page}{2}
\pagestyle{plain}
\baselineskip=18pt
\pagestyle{plain}

\setcounter{footnote}{0}

\tableofcontents

\newpage 
\section{Introduction}

The existence of dark matter (DM) is the clearest hint to physics beyond
the Standard Model (SM). Among various candidates to explain the unknown
component of the Universe, the hypothesis of gravitino dark matter is
very attractive as the gravitino always exists in supersymmetric (SUSY)
theories and is often the lightest superparticle (LSP) since its mass is
suppressed by the Planck scale. The gauge mediated SUSY breaking (GMSB)
scenario~\cite{Dine:1981za,Dine:1993yw} is an explicit realization of
the gravitino LSP while the superpartners of the SM particles can be
much heavier due to the SM gauge interactions.



In the GMSB models, gravitinos are produced 
in the early Universe
from the thermal bath of the
particles in the minimal supersymmetric standard model (MSSM). 
The production process is more effective at high temperatures, 
and thus the relic abundance is proportional to the reheating
temperature after
inflation~\cite{Moroi:1993mb,Kawasaki:1994af,Moroi:1995fs,de
Gouvea:1997tn,Bolz:1998ek,Bolz:2000fu,Pradler:2006qh,Pradler:2006hh,Rychkov:2007uq,Kawasaki:2004qu},
$\Omega_{\rm DM} \propto T_R$.  This gives an upper bound on $T_R$ so as
not for the gravitino abundance to exceed the observed DM abundance,
$\Omega_{\rm DM} h^2 \simeq 0.1$.  The upper bound is $T_R \lesssim 10^6
{\rm GeV}$ for $m_{3/2} \sim 1~{\rm GeV}$ and it becomes more severe for
a lighter gravitino. It is, therefore, difficult to realize the
gravitino DM compatible with the thermal
leptogenesis~\cite{Fukugita:1986hr}, where the maximal baryon asymmetry
is also proportional to $T_R$.
In order to explain the baron asymmetry of the Universe, we need
$T_R \gtrsim 10^9~{\rm GeV}$
\cite{Davidson:2002qv,Giudice:2003jh,Buchmuller:2005eh,Buchmuller:2004nz}.
The ratio $\Omega_{\rm DM} / \Omega_B$ is predicted to be too large
compared to the observed one, i.e., $\Omega_{\rm DM} / \Omega_B \gg 5$.
The late-time entropy production do not help this situation since both
the baryon and DM are diluted while fixing the ratio, $\Omega_{\rm DM} /
\Omega_B$.


The production rate of gravitino has been calculated in the literatures
by using the supergravity Lagrangian, which should be correct at low
energy. However, it has been argued in Ref.~\cite{Choi:1999xm}, those
estimates should be modified in GMSB models for a temperature higher
than the messenger scale $M_{\rm mess}$. The authors of
Ref.~\cite{Choi:1999xm} evaluated the gravitino production rate using
the Lagrangian of global SUSY, and found that for temperature $T \gg
M_{\rm mess}$, the rate is suppressed by $\sim M_{\rm mess}^2 / T^2$
compared to the supergravity calculation. If this is the case, the
gravitino relic abundance becomes insensitive to $T_R$ for $T_R \gg
M_{\rm mess}$. The calculations in global SUSY should match the
supergravity ones for energies higher than the gravitino mass
at the leading order in the $1/M_{\rm pl}$ expansion.

Although the statement in Ref.~\cite{Choi:1999xm} is clear in terms of
global SUSY, it seems somewhat obscure in the supergravity
description.  
In the global SUSY case, the MSSM fields couple to the goldstino (the
longitudinal component of the gravitino) only though the loops of the
messenger fields. The production rate is, therefore, significantly
modified when the energy goes beyond the mass of the messenger fields.
On the other hand, in the supergravity Lagrangian, there are contact
derivative interactions between the gravitino and the supercurrent made
of the MSSM fields, which lead a growth of the amplitude as energy
increases. Therefore, in this description, there is no apparent reason
for the gravitino production to be suppressed above the messenger scale
~\cite{Jedamzik:2005ir, Dalianis:2013pya}.

In this paper, in order to offer a comprehensive view about the
 gravitino thermal production, we explicitly calculate a gravitino
 production process both with a global SUSY Lagrangian and a
 supergravity Lagrangian, independently.  We confirm the suppression of
 the gravitino production rate both in global SUSY and supergravity for
 $\ro{s} > M_{\rm mess}$ even though there is a contact interaction term
 in the supergravity Lagrangian. It is found that the loop diagrams
 involving messenger fields in the supergravity calculation cancel the
 tree-level amplitude at a high energy region. The result 
 agrees with the intuition from the goldstino equivalence.
The results indicate that the relic abundance of the gravitino is
 proportional to the messenger scale, $\Omega_{\rm DM} \propto M_{\rm mess}$
 rather than $T_R$ for $T_R \gg M_{\rm mess}$.  Therefore, in this
 occasion, there is no reason to abandon thermal leptogenesis.  Given
 that the gravitino abundance does not depend on $T_R$, the ratio
 $\Omega_{\rm DM} / \Omega_B$ can be fixed to the observed value, $\sim5$,
 with a suitable $T_R$.  


Although the observed DM-baryon ratio can be explained by the thermal
leptogenesis, the scenario requires a late-time entropy production by
some mechanism, because the produced amount of  gravitino is still
larger than the observation, $\Omega_{\rm DM} h^2 \gg 0.1$,
in order to explain the $\Omega_{\rm DM} / \Omega_B$ ratio. 
Interestingly, we already have a source of the entropy production in
GMSB models; there is a pseudo-moduli field in generic low-energy SUSY
breaking models, which can supply a large amount of entropy by its
decay.
We demonstrate the scenario in a simple model of gauge mediation and
confirm that the scenario indeed works as the mechanism to produce the
right amount of the gravitino DM.

The sketch of the scenario is as follows; the reheating of the Universe
occurs at a high $T_R$ so that the gravitino abundance is independent of
$T_R$.  With an appropriate reheating temperature, the
ratio of energy densities $\Omega_{\rm DM} / \Omega_B$ can be fixed at the
observed value, $\Omega_{\rm DM} / \Omega_B \sim 5$,
 after the reheating process.
Later,
the SUSY breaking pseudo-moduli starts coherent
oscillation about the minimum of the potential, and the oscillation
energy eventually dominates the Universe.  A sizable amount of entropy
is released by the subsequent decay, and the pre-existing gravitinos and
baryons are diluted by a same amount to realize the observed values.

Throughout our analysis, the SUSY scale is assumed to be $M_{\rm SUSY}
\simeq 5~{\rm TeV}$ to realize $m_h = 125~{\rm
GeV}$~\cite{Draper:2011aa} within the MSSM.  Although it sounds
difficult to confirm the scenario by the LHC experiments, the framework
we use predicts a relatively small $\mu$-term and thus 
there is a light
higgsino with $m_{\tilde h} \sim \mathcal{O} (100)~{\rm GeV}$ .  
We explain this point in
appendix~\ref{app:higgsino}.  Such a light higgsino may be within the
reach of future experiments such as at an International Linear Collider
(ILC). Since the life-time of higgsino can be as long as $O(1)$~sec, 
 we check the constraints from
the Big-Bang nucleosynthesis (BBN) and find that the light higgsino is
cosmologically safe if the gravitino mass is less than $\sim 500~{\rm
MeV}$.


\section{Gravitino thermal production in GMSB revisited}

The gravitino production rate has been calculated by using the
supergravity Lagrangian
\cite{Moroi:1993mb,Kawasaki:1994af,Moroi:1995fs,de
Gouvea:1997tn,Bolz:1998ek,Bolz:2000fu,Pradler:2006qh,Pradler:2006hh,Rychkov:2007uq,Kawasaki:2004qu},
which leads the result that the abundance is proportional to $T_R$.
In GMSB models, the production is dominated by that of the longitudinal
mode which can be evaluated by identifying the longitudinal mode as the
goldstino in the global SUSY Lagrangian.
%
%
Moreover, in GMSB models, one can use a framework of a linearly realized
SUSY breaking model with a singlet superfield $S$, whose $F$-component
VEV breaks the SUSY.  

An explicit calculation of the goldstino production shows that the
goldstino relic abundance is not necessarily proportional to
$T_R$~\cite{Choi:1999xm}, which contradicts with the estimation in
supergravity.  We examine this apparent contradiction by calculating the
scattering amplitudes of goldstino/gravitino production process both
with a global SUSY Lagrangian and a supergravity Lagrangian. We confirm
that the supergravity result should be modified at high energy.

\subsection{Gravitino thermal production in GMSB}
{\bf Calculation in supergravity Lagrangian}

Here we briefly review why the gravitino relic abundance is determined
by the reheating temperature $T_R$.  Gravitinos are produced from the
scattering process of the MSSM fields and the amplitudes are calculated
by using the supergravity Lagrangian,
\begin{align}
\mathcal{L}_{\rm sugra}^{\rm MSSM}
	\ni
	-\frac{1}{\ro{2} M_{\rm pl}} 
	(D_\nu \phi_i)^\ast \bar{\psi}_{3/2 \mu}  \gamma^\nu \gamma^\mu
	P_L \psi_i
	-\frac{i}{4 M_{\rm pl}} \bar{\lambda}^a \gamma^\mu 
	[ \gamma^\nu , \gamma^\rho ]
	\psi_{3/2 \mu} F_{\nu \rho}^a
	+{\rm h.c.}	
		,
	\label{Lsugra}
\end{align}
where the gravitino field is denoted by $\psi_{3/2 \mu}$.  The gravitino
has the tree-level interactions with all the chiral multiplets ($\phi_i,
\psi_i$) or gauge multiplets ($A_\mu^a, \lambda^a$) in the MSSM and the
form of interactions is uniquely fixed by local SUSY.

\begin{figure}[t]
\begin{center}
\includegraphics[width=5cm]{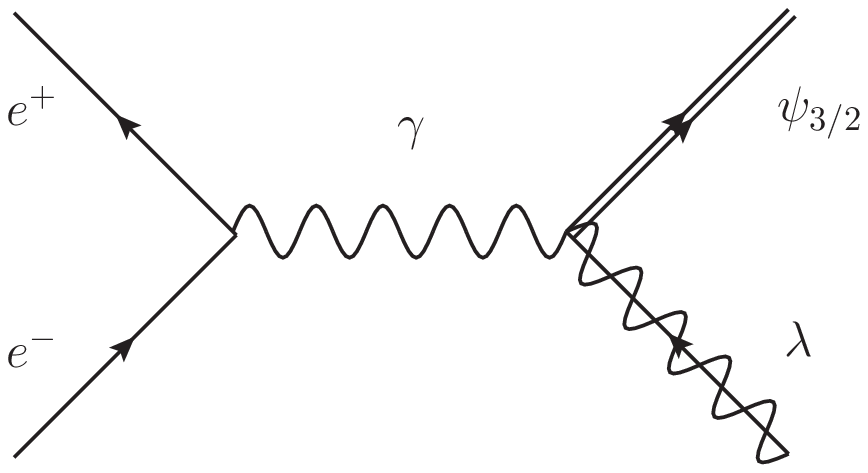}
\includegraphics[width=4.5cm]{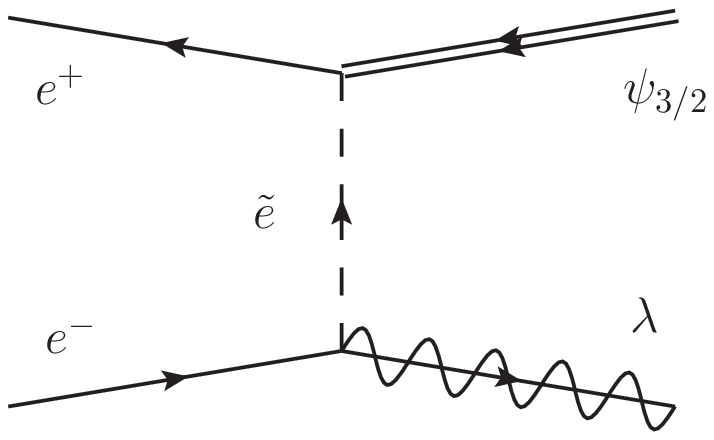}
\includegraphics[width=5cm]{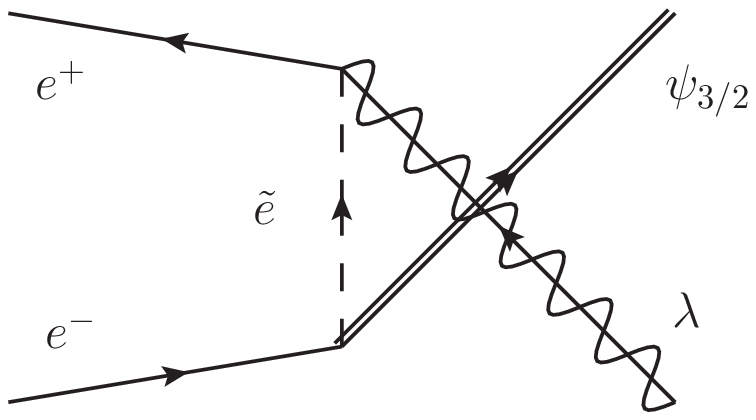}
\caption{Gravitino production process
	$e^- e^+ \to \lambda \psi_{3/2}$.
		 }
\label{tree}
\end{center}
\end{figure}

For the gravitino production, there are ten two-body processes involving
left-handed quarks ($q_i$), squarks ($\tilde q_i$), gauginos
($\lambda^a$) and the gauge bosons ($A^a$), which are called processes A
to J in the literatures
~\cite{Moroi:1993mb,Kawasaki:1994af,Moroi:1995fs,Bolz:1998ek,Bolz:2000fu}. 
In the literatures the QCD
processes are discussed in detail because they are the dominant
processes.  
Here we focus on a particular process $e^- e^+ \to \lambda \psi_{3/2}$
(called process I in the literatures) for simplicity.
The tree-level diagrams are shown in Fig.~\ref{tree}.

The scattering amplitude is calculated by the supergravity Lagrangian in
Eq.~(\ref{Lsugra}).  Among the polarized amplitudes, the following turns
out to have the highest power in the center-of-mass energy, $\sqrt s$,
and thus dominates at high energies,
\begin{align}
	\mathcal{M}_{e^- e^+ \to \lambda \psi_{3/2}}^{(\uparrow \downarrow \uparrow \uparrow)}
	=
	\frac{e m_\lambda}{\ro{6} m_{3/2} M_{\rm pl}} 
	\ro{s}
	\sin \theta,
	\label{amp1}
\end{align}
where arrows in the parenthesis represent the spins of the electron, the
positron, the gaugino and the gravitino, respectively.  
The angle $\theta$ is the production angle in the center-of-mass frame.
The gauge coupling of QED is denoted by $e$.  Although each of $s$-,
$t$- and $u$-channel diagrams has an energy dependence of $O(s)$, they
are canceled out when combined, remaining the energy dependence of
$O( \ro{s} )$.  
The above contribution is from the longitudinal component of the gravitino 
whose wave function is approximately proportional to $\sqrt s / m_{3/2}$ with $m_{3/2}$ the gravitino mass.

In order to estimate the relic abundance of the gravitino,
we should calculate the reaction rate which is proportional to the square of the amplitude,
\begin{align}
	\Gamma_{e^- e^+ \to \lambda \psi_{3/2}} (T)
	\propto
	\frac{m_{\lambda}^2}{m_{3/2}^2 M_{\rm pl}^2}
	T^3,
	\label{gamma1}
\end{align}
where the temperature dependence is determined by dimensional analysis.
The key is the cubic dependence on $T$. 
If the reaction rate depends on the temperature with a higher power than the
Hubble parameter $H(T) \propto T^2$, the resultant gravitino abundance
is fixed at high temperature, $T_R$.  In contrast, if the power is lower
than $H(T)$, the yield is fixed by the lowest temperature.  If the
process $e^- e^+ \to \lambda \psi_{3/2}$ is effective and
Eq.~(\ref{gamma1}) is valid for an arbitrary temperature, the gravitino
abundance is determined by $T_R$.  

\noindent
{\bf Goldstino analysis}

\begin{figure}[t]
\begin{center}
\includegraphics[width=7.4cm]{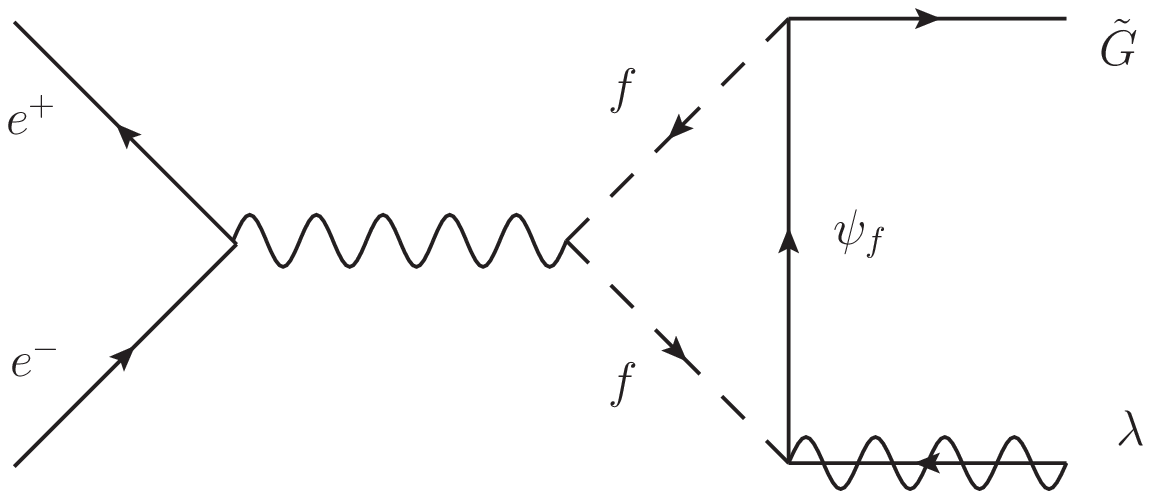}
\includegraphics[width=7.4cm]{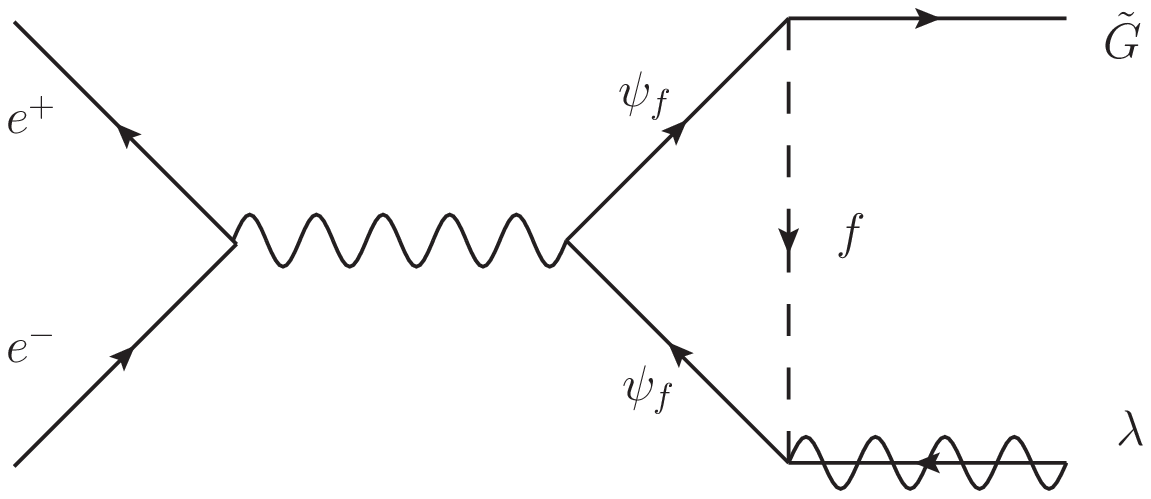}
\caption{Goldstino production process
	$e^- e^+ \to \lambda \tilde{G}$.
		 }
\label{goldstino}
\end{center}
\end{figure}

In GMSB models, effects of SUSY breaking are transmitted to the MSSM sector through the messenger 
loop diagrams.
A superpotential of the following form is usually assumed,
\begin{align}
	W= \lambda S f \bar{f}.
	\label{W1}
\end{align}
SUSY is broken by the $F$-component of  the singlet superfield $S$.
$f$ and $\bar f$ represent the messenger superfields which have SM gauge charges.
If $F_S$ is the only source of the SUSY breaking, 
the fermion component of $S$ (we call it $\psi_S$) is  the goldstino $\tilde{G}$,
 which is absorbed into the longitudinal component of the gravitino.
In general, there are additional sources of SUSY breaking from the $F$-components of other chiral multiplets.
In that case, the goldstino is composed of the liner combination of the fermions which belong to the multiplets
whose $F$-components develop VEVs,
\begin{align}
	\tilde{G} =
	\frac{F_S}{F} \psi_S +  \sum_i \frac{F_i}{F} \psi_i,
\end{align}
where $F= \ro{|F_S |^2 + \sum_i |F_i |^2}$.
Therefore,
the amplitude for the goldstino production is given by rescaling that for $\psi_S$ 
by a factor $F_S / F$.
Unlike the gravitino in the supergravity Lagrangian,
 the goldstino does not couple directly to the MSSM fields.
The goldstino is produced  through the messenger loop diagrams shown in Fig.~\ref{goldstino}.
We expect that the scattering amplitude of the process $e^- e^+ \to \lambda \tilde{G}$ 
coincides that of the gravitino production in Eq.~(\ref{amp1}).

By explicitly evaluating these diagrams, however, a different result
from supergravity estimation comes out.  For the same process and the
same polarization to Eq.~(\ref{amp1}), the scattering amplitude is
calculated to be
\begin{align}
	\mathcal{M}_{e^- e^+ \to \lambda \tilde{G}}^{(\uparrow \downarrow \uparrow \uparrow)}
	&=
		-  \frac{2 \ro{2} e^3 \lambda}{(4 \pi )^2}
		\frac{F_S}{F}
		 M_{\rm mess}
		C_0 (\ro{s}, M_{\rm mess}) \ro{s} \sin \theta
	\\
	&=
	- \frac{2e m_\lambda M_{\rm mess}^2}{\ro{6} m_{3/2} M_{\rm pl}}
	C_0 (\ro{s}, M_{\rm mess}) \ro{s}
	\sin \theta,
	\label{amp2}
\end{align}
where $M_{\rm mess} = \lambda \langle S \rangle$ is the messenger mass scale.
We have translated the parameters of global SUSY, $\lambda$ and $\langle S \rangle$,
to the parameters of the supergravity, $m_{3/2}$ and $M_{\rm pl}$
by using the formulae in GMSB:
\begin{align}
	m_\lambda = \frac{2 e^2}{(4 \pi)^2}
		\frac{F_S}{\langle S \rangle},
	\label{gauginomass}
\end{align}
and 
\begin{align}
	m_{3/2} =
	\frac{F}{\ro{3} M_{\rm pl}}.
\end{align}

The function $C_0 (\ro{s}, M_{\rm mess})$ is the $C$-function defined in
Ref.~\cite{'tHooft:1978xw},
\begin{align}
	C_0 (\ro{s}, M_{\rm mess})
	&=
	\int_0^1 dx \frac{1}{s(1-x)} \log
	\left[
		1-\frac{s}{M_{\rm mess}^2} x(1-x) -i \epsilon
	\right].
	\label{C0}
\end{align}
In a low energy limit, $\sqrt s \ll M_{\rm mess}$, $C_0$ is
approximately given by $C_0 \simeq -1/2 M_{\rm mess}^2$ and
reproduces the result of supergravity calculation in Eq.~(\ref{amp1}).
However, for $\ro{s} \gg M_{\rm mess}$, $C_0$ scales as $1/s$ up to
a logarithmic factor.

If the external energies are lower than the messenger mass scale, i.e., for
 $T < M_{\rm mess}$, the reaction rate
depends on the temperature as $\propto T^3$,
\begin{align}
	\Gamma_{e^- e^+ \to \lambda \tilde{G}} (T)
	\propto
	\frac{m_\lambda^2}{m_{3/2}^2 M_{\rm pl}^2}
	T^3,
	\ \ \ \ \ \ 
	{\rm for}
	\ \  
	 T \ll M_{\rm mess} ,
	 \label{gammas1}
\end{align}
which reproduces the result of the supergravity calculation in Eq.~(\ref{gamma1}).
Here we again squared the amplitude and fixed the temperature dependence
by dimensional analysis.  However, for $T> M_{\rm mess}$, the reaction
rate is suppressed by $\sim M_{\rm mess}^2~/~T^2$ compared
to Eq.~(\ref{gammas1}), namely
\begin{align}
	\Gamma_{e^- e^+ \to \lambda \tilde{G}} (T)
	\propto
	\frac{m_\lambda^2 M_{\rm mess}^2}{m_{3/2}^2 M_{\rm pl}^2} T
	,
	\ \ \ \ \ \ 
	{\rm for}
	\ \  
	 T \gg M_{\rm mess} .
	 \label{gammas2}
\end{align}
The point is that the temperature dependence of $\Gamma_{e^- e^+ \to
\lambda \tilde{G}}(T) / H(T)$ gets suppressed as $1/T$ at high
temperatures, which makes the goldstino relic abundance irrelevant to
the reheating temperature.  Rather, the abundance is determined by the
messenger mass scale.  


\newpage
\noindent
{\bf Supergravity calculation in GMSB}

We observe a difference between the two amplitudes, Eq.~(\ref{amp1}) and
Eq.~(\ref{amp2}).  One of them should be modified at high energy,
$\ro{s} \gg M_{\rm mess}$, if we believe in the goldstino
equivalence.

\begin{figure}[t]
\begin{center}
\includegraphics[width=7cm]{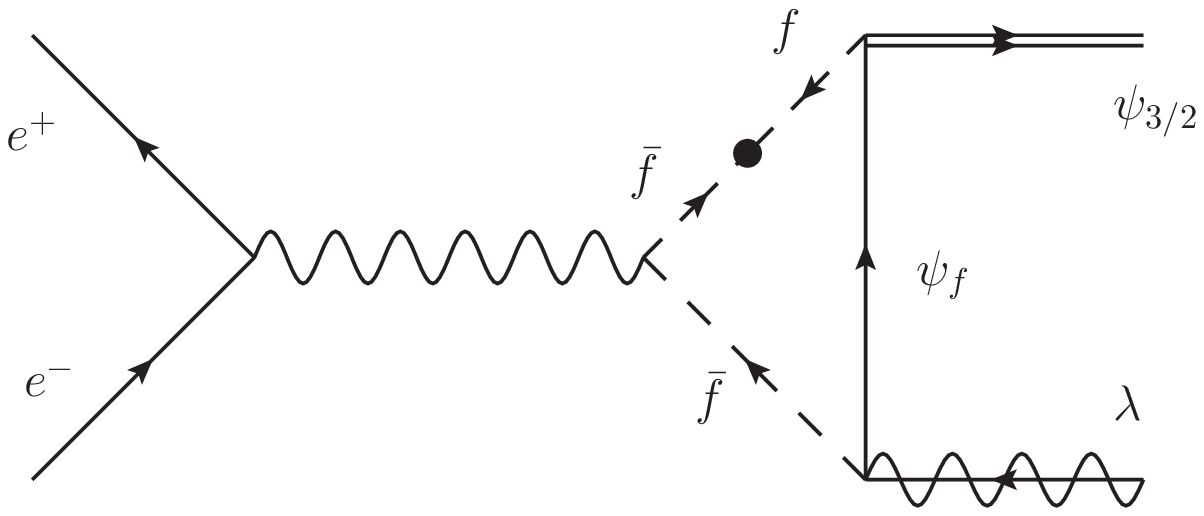}
\includegraphics[width=7cm]{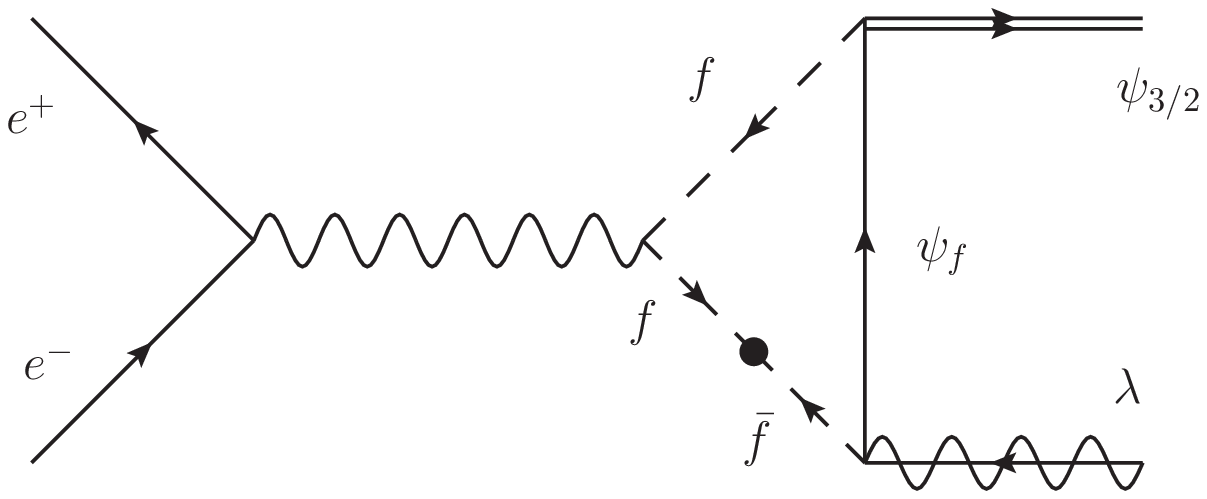}

\includegraphics[width=7cm]{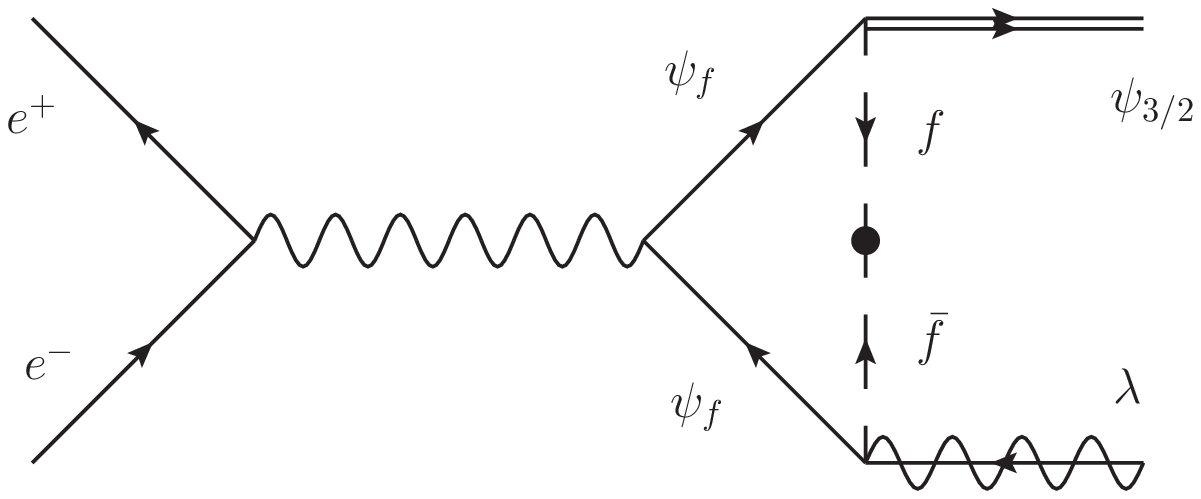}

\includegraphics[width=6cm]{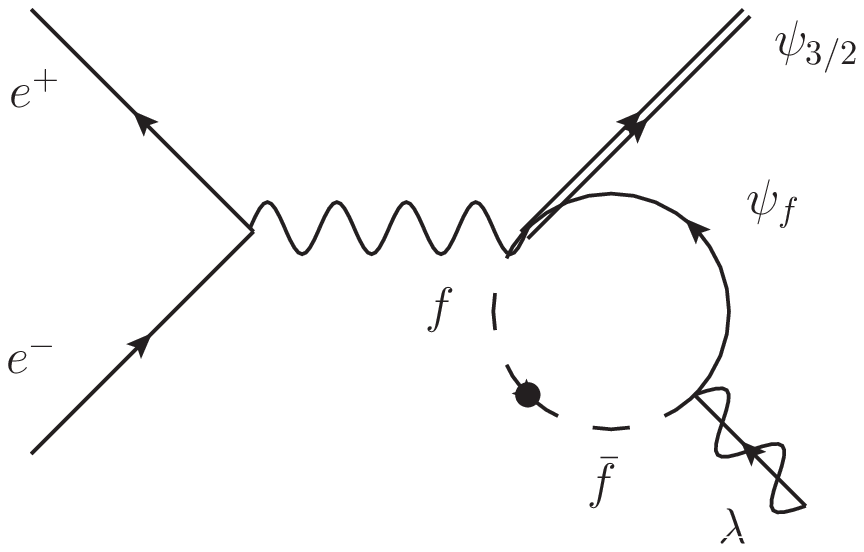}
\caption{One-loop diagrams for the gravitino production
	$e^- e^+ \to \lambda \psi_{3/2}$.
		 }
\label{messengerloop}
\end{center}
\end{figure}

We find that the modification appears in the supergravity calculation.
In GMSB models, there are messenger fields, which potentially affect the
gravitino production process.  In fact, they contribute to the gravitino
production process $e^- e^+ \to \lambda \psi_{3/2}$ through the one-loop
diagrams shown in Fig.~\ref{messengerloop}. Even though they are
diagrams at the one-loop level, they cannot be neglected compared to the
tree-level ones in Fig.~\ref{tree} since the gaugino mass in
Eq.~\eqref{amp1} is at the one-loop order in GMSB models.
Note here that the diagrams in Fig.~\ref{messengerloop} are not 
the microscopic description of the first diagram in Fig.~\ref{tree}.
Both diagrams exist as independent ones in supergravity.
%
The explicit calculation shows
\begin{align}
	\mathcal{M}_{e^- e^+ \to \lambda \psi_{3/2}}^{(\uparrow \downarrow \uparrow \uparrow) }
	({\rm one~loop})
	=
	-
	\frac{e m_\lambda}{\ro{6} m_{3/2} M_{\rm pl}} 
	\ro{s}
	\sin \theta
	\left[
	2 M_{\rm mess}^2 C_0 (\ro{s} , M_{\rm mess}) +1
	\right],
	\label{amp3}
\end{align}
where $C_0$ is again the $C$-function in Eq.~(\ref{C0}).
The dots in Fig.~\ref{messengerloop} represent insertions of $F_S$,
and we used Eq.~(\ref{gauginomass}) to derive the above formula.
A few comments
are in order.  At a lower energy than the messenger mass scale, the
messenger fields can be integrated out and absent in the low energy theory. 
The gravitino interactions are then completely read off from the
supergravity Lagrangian of the MSSM fields (\ref{Lsugra}). The
supergravity prediction in Eq.~(\ref{amp1}), therefore, should not be
altered for $\ro{s} \ll M_{\rm mess}$.  
The additional contribution (\ref{amp3}) indeed respects this
consideration.  The factor, $2M_{\rm mess}^2 C_0 +1 $, in
Eq.~(\ref{amp3}) goes to zero as $\ro{s} \to 0$, and thus the amplitude
is accurately represented by Eq.~(\ref{amp1}) at  low energy.  However,
the one-loop contribution becomes comparable to that of tree-level for
$\ro{s} \gg M_{\rm mess}$ since the factor, $2M_{\rm mess}^2 C_0 +1$,
approaches to $1$.

Combined with the tree-level contribution ({\ref{amp1}), we confirmed
that the growing amplitude at $\sqrt s \gg M_{\rm mess}$ in supergravity
is completely cancelled by the one-loop diagrams, and the total
supergravity calculation coincides with the result from global SUSY,
\begin{align}
	\mathcal{M}_{e^- e^+ \to \lambda \psi_{3/2}}^{(\uparrow \downarrow \uparrow \uparrow) }
	&=
	\mathcal{M}_{e^- e^+ \to \lambda \psi_{3/2}}^{(\uparrow \downarrow \uparrow \uparrow) } ({\rm tree})
	+
	\mathcal{M}_{e^- e^+ \to \lambda \psi_{3/2}}^{(\uparrow \downarrow \uparrow \uparrow) } ({\rm one~loop}) \notag \\
	&=
	\frac{e m_\lambda}{\ro{6} m_{3/2} M_{\rm pl}} 
	\ro{s}
	\sin \theta
	-
	\frac{e m_\lambda}{\ro{6} m_{3/2} M_{\rm pl}} 
	\ro{s}
	\sin \theta
	\left[
	2 M_{\rm mess}^2 C_0 (\sqrt s, M_{\rm mess}) +1
	\right] \notag \\
	&=
	- \frac{2e m_\lambda M_{\rm mess}^2}{\ro{6} m_{3/2} M_{\rm pl}}
	C_0 (\sqrt s, M_{\rm mess}) \ro{s}
	\sin \theta.
\end{align}


\noindent
{\bf Additional contribution from the tree-level messenger scatterings}

For $T>M_{\rm mess}$, in addition to the scattering processes of the MSSM
particles, the goldstino is also produced by scattering processes where
the messenger fields are in the external lines.  The reaction rate is
calculated to be~\cite{Choi:1999xm}
\begin{align}
	\Gamma_{{\rm messengers} \to \lambda \tilde{G}} (T) 
&\propto
	\lambda^2 \left( \frac{F_S}{F} \right)^2 T \notag \\
&\propto
	\left(
		\frac{4 \pi}{\alpha}
	\right)^2
	\frac{m_\lambda^2 M_{\rm mess}^2}{m_{3/2}^2 M_{\rm pl}^2} T
	.
	\label{gammas3}
\end{align}
As we see from Eq.~(\ref{gammas2}) and Eq.~(\ref{gammas3}), the reaction
rate of the messenger particles is larger than that of the MSSM
particles by a loop-factor since the messenger fields directly couple to
the goldstino through the superpotential interaction.

\subsection{The gravitino relic abundance}

Summarizing the previous subsection, in GMSB models, the gravitino is
 produced from the scattering processes of the MSSM fields and the
 messenger fields.  Depending on the value of $T_R$, the resultant
 gravitino relic abundance is determined by different values; if
 $T_R<M_{\rm mess}$, the abundance is fixed by $T_R$, and if $T_R>M_{\rm
 mess}$, it is the messenger mass scale to fix the abundance,
\begin{align}
\Omega_{3/2} h^2
	\simeq 0.45 \left( \frac{T_R}{10^6 {\rm GeV}} \right)
		\left( \frac{\rm GeV}{m_{3/2}} \right)
		\left( \frac{m_{\tilde g}}{5~{\rm TeV}} \right)^2 
		\ \ \ (T_R < M_{\rm mess}),
		\label{omega321}
		\\
\Omega_{3/2} h^2
	\simeq 3.7 \times 10^2 \left( \frac{M_{\rm mess}}{10^6 {\rm GeV}} \right)
		\left( \frac{\rm GeV}{m_{3/2}} \right)
		\left( \frac{m_{\tilde g}}{5~ {\rm TeV}} \right)^2
		\ \ \ (T_R > M_{\rm mess}).
		\label{omega322}
\end{align}

The abundance in Eq.~(\ref{omega322}) is not a straightforward replacement
 of $T_R$ to $M_{\rm mess}$ in Eq.~(\ref{omega321}) since the production
 through the messenger fields are not suppressed by a loop factor.


\begin{figure}[t]
\begin{center}
\includegraphics[width=12cm]{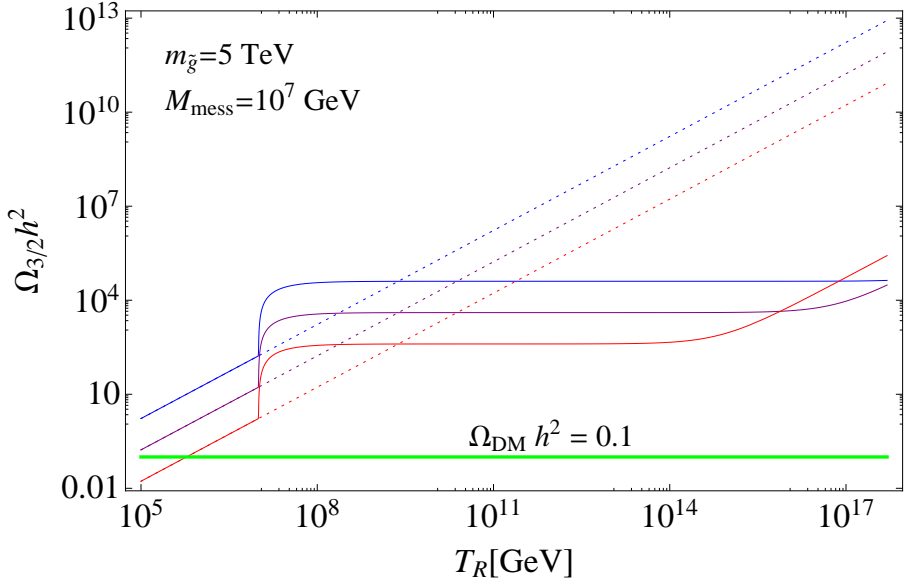}
\caption{Gravitino relic abundance. Blue, purple, and red lines represent $m_{3/2} =100~{\rm MeV}$,
 $m_{3/2} =1~{\rm GeV}$ and $m_{3/2} = 10~{\rm GeV}$, respectively.
The gravitino abundance become insensitive to the reheating temperature for $M_{\rm mess} < T_R$
(solid lines).
Dotted lines are naive extrapolations of Eq.~(\ref{omega321}).
For a very high reheating temperature ($T_R \gtrsim 10^{14}~{\rm GeV}$),
the transverse mode of the gravitino becomes important.
}
\label{omega32}
\end{center}
\end{figure}

The estimates so far do not include a contribution of the transverse
mode of the gravitino.  For a very high reheating temperature, the
transverse mode becomes relevant,
\begin{align}
\Omega_{3/2} h^2 ({\rm transverse})
	\simeq 0.53 \left( \frac{T_R}{10^{13} {\rm GeV}} \right)
		\left( \frac{m_{3/2}}{\rm GeV} \right).
\end{align}

Including both the longitudinal and the transverse modes, we show the
gravitino relic abundance in GMSB with the messenger scale fixed to be
$M_{\rm mess} = 10^7~{\rm GeV}$ in Fig.~\ref{omega32}.  As we see from
the figure, the gravitino relic abundance is predicted to be constant in
a wide range of the reheating temperature, but the amount is too large
compared to the observed dark matter energy density $\Omega_{\rm DM} h^2
\simeq 0.1$.  The overproduced gravitinos must be diluted by some
mechanism.  Although the prediction to $\Omega_{3/2}$ is too large, the
insensitivity to $T_R$ brings us a new scenario of gravitino DM.

\section{A new scenario of gravitino dark matter}

As we have confirmed in the previous section, the gravitino relic
abundance becomes insensitive to $T_R$ once the temperature of the
Universe exceeds the messenger mass scale.  The results have a crucial
impact on the possible mechanism of baryogenesis.  In this section, we
present a new cosmological scenario of gauge mediation, where gravitino
dark matter and thermal leptogenesis are compatible.  The scenario
requires a late-time entropy release by some mechanism, which is
 automatically supplied by the decay of the SUSY breaking
pseudo-moduli field.  We demonstrate the scenario with a simple model of
gauge mediation as an example and see that the scenario actually works.

Throughout the analysis, we assume the SUSY scale to be $M_{\rm SUSY}
\gtrsim 5$ TeV and in particular fix the gluino mass to be $m_{\tilde g}
= 5$ TeV to account for the Higgs boson mass of 125 GeV within the MSSM
in GMSB~\cite{Draper:2011aa}.  Although most of the SUSY particles are
then too heavy to be detected at the LHC experiments, the model predicts
higgsino to be as light as $m_{\tilde h} \sim \mathcal{O} (100)~{\rm
GeV}$.  We briefly mention the cosmological constraint on the light
higgsino in the last subsection.

\begin{figure}[t]
\begin{center}
\includegraphics[width=12cm]{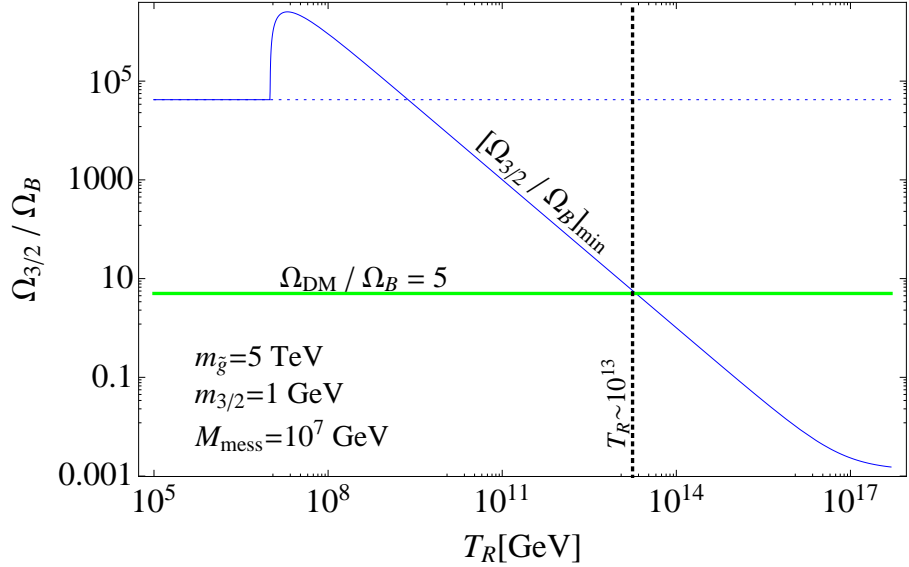}
\caption{Since the gravitino abundance becomes constant for $M_{\rm mess} < T_R$ whereas 
the maximum value of $\Omega_B$ is 
always proportional to $T_R$, the ratio $\Omega_{3/2} / \Omega_B$ eventually reaches the observed value
as $T_R$ becomes higher. 
We plotted a minimum value of the prediction for $\Omega_{3/2} / \Omega_B$ as a function of $T_R$.
We see that the observed value of $\Omega_{\rm  DM} / \Omega_B$
can be reproduced for $T_R \gtrsim 10^{13} {\rm GeV}$.
		 }
\label{compatibility}
\end{center}
\end{figure}

\subsection{Compatibility with thermal leptogenesis}

In a light gravitino scenario, 
thermal leptogenesis and gravitino DM are thought to be incompatible with each other.
The possible maximum amount of baryon asymmetry produced by the thermal leptogenesis is 
proportional to the reheating temperature~\cite{Davidson:2002qv,Giudice:2003jh,Buchmuller:2005eh,Buchmuller:2004nz},
\begin{align}
	\Omega_B
	\lesssim
	0.04
	\left(
		\frac{T_R}{10^{9} {\rm GeV}}
	\right),
\end{align}
which puts a lower bound on $T_R$ ($T_R \gtrsim 10^9 {\rm GeV})$ to
 realize the observed value $\Omega_B \simeq 0.045$.  If the gravitino
 relic abundance is represented as Eq.~(\ref{omega321}) for any $T_R$,
 the thermal production of gravitino DM and the thermal leptogenesis are incompatible; even if
 we assume a late-time entropy production to dilute overproduced
 gravitino to match the abundance to the observation, baryons are also
 diluted at the same time and the abundance never reproduces the
 observation.  In other words, the ratio $\Omega_{3/2} / \Omega_B$ is
 constant as long as the abundances are both proportional to $T_R$, and
 always larger than the observed ratio, $\Omega_{\rm DM} / \Omega_B \sim
 5$.

However, in GMSB, if the reheating temperature is higher than the
messenger mass scale, the gravitino relic abundance becomes insensitive
to $T_R$.  Then, the observed ratio of the energy densities, $\Omega_{\rm
DM} / \Omega_B \sim 5$, can be realized with thermally produced gravitino
and the thermal leptogenesis.  We plot the prediction for $\Omega_{3/2}
/ \Omega_B$ to visualize the situation in Fig.~\ref{compatibility}.  If
the gravitino abundance is proportional to $T_R$ for any $T_R$, the
theoretical prediction never reaches the observed value $\Omega_{\rm DM}
/ \Omega_B \sim 5$ (dotted line).  However, if the reheating temperature
is higher than the messenger scale, $\Omega_{3/2}$ becomes independent
of $T_R$ in GMSB, which allows $\Omega_{3/2} / \Omega_B$ to achieve the
observed value.  

\subsection{Late-time entropy release}

The ratio of the energy densities  $\Omega_{\rm DM} / \Omega_B \sim 5$
can be realized by thermally produced gravitino and thermal leptogenesis 
with an appropriate reheating temperature as we saw above.
However, as is obvious from Fig.~\ref{omega32},
the predicted gravitino abundance is too large compared to the observation, $\Omega_{\rm DM} h^2 \simeq 0.1$.
The overproduced gravitino should be diluted by a late-time entropy release 
by some mechanism.
The required amount of dilution is 
\begin{align}
	\Delta_{3/2}
	&\equiv
	\frac{\Omega_{3/2} h^2}{\Omega_{\rm DM} h^2} \\
	&\simeq
	7.5 \times 10^4 
	\left(
	\frac{M_{\rm mess}}{10^7~{\rm GeV}}
	\right)
	\left(
	\frac{500~{\rm MeV}}{m_{3/2}}
	\right)
	\left(
	\frac{m_{\tilde g}}{5~{\rm TeV}}
	\right)^2,
	\label{delta32}
\end{align}
where $T_R > M_{\rm mess}$ is assumed.

Actually, a source of entropy production is already incorporated in the
scenario: the scalar component of the singlet superfield $S$, which is
called the pseudo-moduli field.  In the early Universe, it is possible
that the pseudo-moduli is displaced from the vacuum and
starts oscillation around the minimum.  Since the pseudo-moduli is
massless at tree-level and gets mass only through the quantum effects,
it is often much lighter than the SUSY breaking scale, $\sqrt F$, and is
long-lived if there is a weakly coupled description for the SUSY
breaking sector.  In such a case, the pseudo-moduli can eventually
dominate the energy density of the Universe, and a sizable amount of
entropy is produced from its decay.


\subsection{Demonstration in a simple model of gauge mediation}

\noindent
{\bf The model}

We study a low-energy effective theory of O'Raifeartaigh type SUSY
breaking model coupled with the messenger fields:
\begin{align}
	&K=
		f^\dagger f + \bar{f}^\dagger \bar{f} + S^\dagger S - \frac{(S^\dagger S )^2}{\Lambda^2}
		+ \cdots, 
		\label{kahler}
		\\
	&W=
		m^2 S - \lambda S f \bar{f}  + c,
		\label{super}
\end{align}
where $S$ is a gauge singlet superfield called the SUSY breaking pseudo-moduli.
The messenger superfields are demoted by $f$ and $\bar{f}$.
There is an $R$-symmetry where the charge assignment is $R(S)=2$ and
$R(f \bar{f})=0$.  If the $R$-symmetry is unbroken, $S$ is stabilized at
$S=0$ where we cannot integrate out the messenger fields.  Once we turn
on the supergravity effects, however, the $R$-symmetry is explicitly
broken by the supergravity correction represented by the constant term
$c$, which destabilizes the origin and creates the SUSY breaking vacuum
at $\langle S \rangle \sim \Lambda^2 / M_{\rm pl}$~\cite{Kitano:2006wz},
where $M_{\rm pl} \simeq 2.4 \times 10^{18} {\rm GeV}$ is the reduced
Planck scale.

Since there is also a SUSY preserving minimum at $S=0$ where the
messenger fields condense, the SUSY breaking vacuum is a meta-stable
state. For a realistic cosmology, the $S$ field should stay away from
the SUSY vacuum in the course of cosmological evolution.

%

\newpage
\noindent
{\bf Cosmological evolution of $S$}

Cosmological evolution of the pseudo-moduli in the model in
Eqs.~(\ref{kahler}) and (\ref{super}) is examined in detail in
Refs.~\cite{Ibe:2006rc,Hamaguchi:2009hy,Fukushima:2012ra,Dalianis:2010yk,Dalianis:2010pq}.
It has been found that the SUSY breaking minimum is preferred to the
SUSY preserving one for a wide region of the parameter space even if the
messenger particles enter the thermal equilibrium\footnote[3]{The vacuum
selection is discussed in the literatures
\cite{Abel:2006cr,Craig:2006kx,Fischler:2006xh,Abel:2006my,Anguelova:2007at,
Papineau:2008xf,Auzzi:2010wm,Katz:2009gh,Moreno:2009nk,Ferrantelli:2009zv}.
The present model had been thought to be problematic because 
the $S$ field tends to fall into the SUSY preserving vacuum by a
finite temperature potential.  }.
Also, with an appropriate initial condition the pseudo-moduli start
oscillation around the SUSY breaking vacuum and the oscillation energy
dominates the energy density of the Universe. 

We define the dilution factor $\Delta$ due to the entropy release from
the decay of the pseudo-moduli as
\begin{align}
	\frac{1}{\Delta}
	\equiv
	\frac{s_{\rm inf}}{s_S + s_{\rm inf}}
	\simeq
	{\rm Min} \left[ 1, \frac{s_{\rm inf}}{s_S} \right],
\end{align}
where $s_{\rm inf}$ and $s_S$ represent the entropy densities produced by the decays of the inflaton and 
$S$, respectively.
If $\Delta > 1$, $\Delta$ is well approximated by
\begin{align}
	\Delta \simeq
	\frac{s_S}{s_{\rm inf}}
	=
	\frac{4}{3 T_d} \cdot \frac{\rho_S}{s_{\rm inf}},
\end{align}
where $\rho_S$ is the energy density of $S$ and 
$T_d$ is a decay temperature of the pseudo-moduli, which is defined by
\begin{align}
	T_d
	\equiv
	\left(
		\frac{\pi^2 g_\ast}{90}
	\right)^{-1/4}
	\ro{M_{\rm pl} \Gamma_S}.
	\label{Td}
\end{align}
The  total decay width of $S$ is denoted as $\Gamma_S$.
The formulae of $\Gamma_S$ and $T_d$ are found in appendix~\ref{app:int}.

If the magnitude of dilution factor $\Delta$ coincides $\Delta_{3/2}$ in
Eq.~(\ref{delta32}), the overproduced gravitinos are diluted to realize
the observed dark matter abundance, $\Omega_{\rm DM} h^2 \simeq 0.1$.  In
order to realize the right amount of baryons, $\Omega_B \simeq 0.045$,
at the same time, we need an appropriate reheating temperature.  Since
the baryon asymmetry is also diluted by the entropy
production, the reheating temperature should be high enough to produce
abundant baryons in advance, namely $10^9 \times \Delta_{3/2} \lesssim
T_R$ is required in the scenario.  We show the required set of the
dilution factor ($\Delta_{3/2}$) and the reheating temperature ($T_R$)
in $m_{3/2}$ vs $M_{\rm mess}$ plane in Fig.~\ref{delta_TR}.

In the present set-up,
there exists
 a parameter region where the dark matter and the baryon asymmetry are
 explained by thermally produced gravitino and thermal leptogenesis
 simultaneously (blue and green regions), with an appropriate combination of
 $\Delta$ and $T_R$. 

In order to estimate the magnitude of the dilution factor from the
decay, we numerically solved the equation of motion of the pseudo-moduli
with the initial condition set at the inflaton dominated era.  The
results depend on the initial location of the $S$ field which can be far
away from the origin depending on the inflation model and the coupling
between $S$ and the inflaton~\cite{Ibe:2006rc,Hamaguchi:2009hy}. In this
study, we choose the initial position of $S$ to be $\Lambda$ or $M_{\rm
pl}$ for illustration.  The results are shown in
Fig.~\ref{dilution_result}.  As we see from the figure, by choosing an
appropriate value of the initial condition of $S$ from between $\Lambda$
and $M_{\rm pl}$, the required amount of entropy can be supplied from
the oscillation energy everywhere in the blue and green regions in
Fig.~\ref{delta_TR}; we have confirmed that required entropy production
can be obtained in this model.

\noindent
{\bf Non-thermal gravitino production}

While the dark matter is explained by thermally produced gravitino in
the blue and green regions in Fig.~\ref{delta_TR}, gravitinos are also
produced non-thermally by the rare decay $S \to \psi_{3/2} \psi_{3/2}$.
We calculate the non-thermally produced gravitino abundance in
appendix~\ref{app:int} and found that the abundance coincides the observed
dark matter abundance with $m_{3/2} \sim 2~{\rm GeV}$.  Taking into
account possible theoretical errors, we show the parameter region where
$0.03 \lesssim \Omega_{3/2}^{\rm NT} h^2 \lesssim 0.3$ is predicted as a
green band in Fig.~\ref{delta_TR}.

\begin{figure}[h]
\begin{center}
\includegraphics[width=11cm]{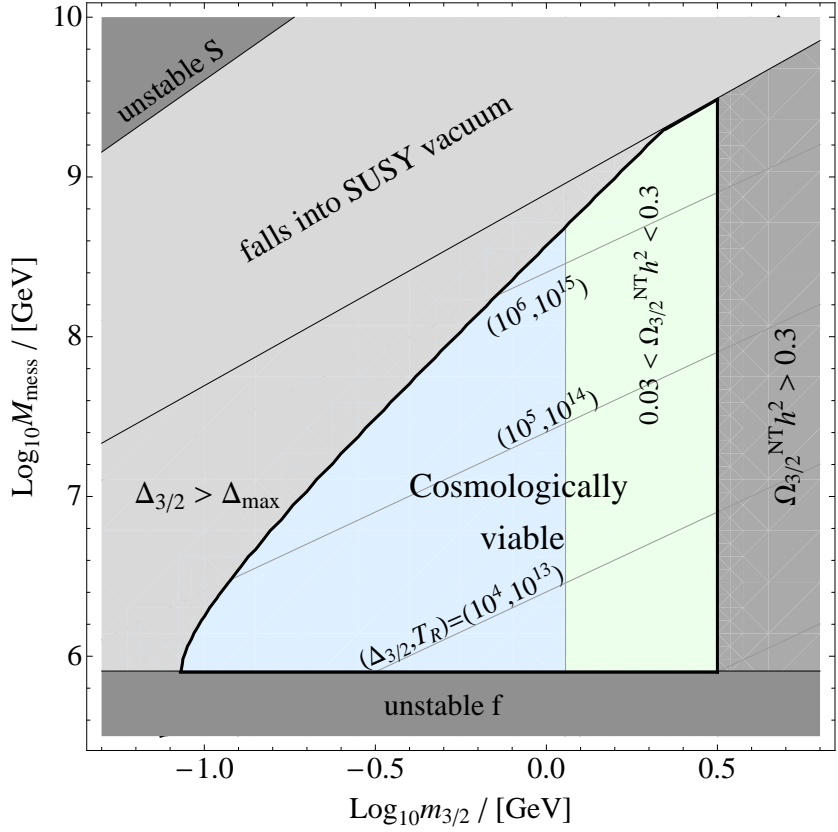} \caption{Required amount of
the dilution factor ($\Delta_{3/2}$) and the reheating temperature
($T_R$) to realize the observation $\Omega_{\rm DM} h^2 \simeq 0.1$ and
$\Omega_B \simeq 0.045$.  In blue and green regions, the dark matter is
explained by gravitino and baryon asymmetry is supplied by thermal
leptogenesis with an appropriate choice of $\Delta$ and $T_R$.  In the
green region, the non-thermally produced gravitino abundance coincides
the observed DM abundance.  We should discard the parameter regions
shaded by (light)gray color.  For gray regions denoted as ``unstable
$S$'' and ``unstable $f$,'' the SUSY breaking minimum is
unstable~\cite{Kitano:2006wz}.  For a light gray region ``fall into SUSY
vacuum,'' the pseudo-moduli fall into SUSY preserving vacuum along the
cosmological evolution and never reaches the SUSY breaking
vacuum~\cite{Fukushima:2012ra}.  We define $\Delta_{\rm max}$ as the
maximum dilution factor available under the condition that the
oscillation amplitude is small so that $S$ does not fall into SUSY
vacuum.  In the region $\Delta_{3/2} > \Delta_{\rm max}$ we cannot
obtain a required amount of dilution factor $\Delta_{3/2}$ while $S$
successfully reaches the SUSY breaking minimum.  Gravitinos are
overproduced non-thermally in the gray region ``$\Omega_{3/2}^{\rm NT}
h^2 >0.3$.''  } \label{delta_TR}
\end{center}
\end{figure}
\clearpage

\begin{figure}[t]
\begin{center}
\includegraphics[width=7cm]{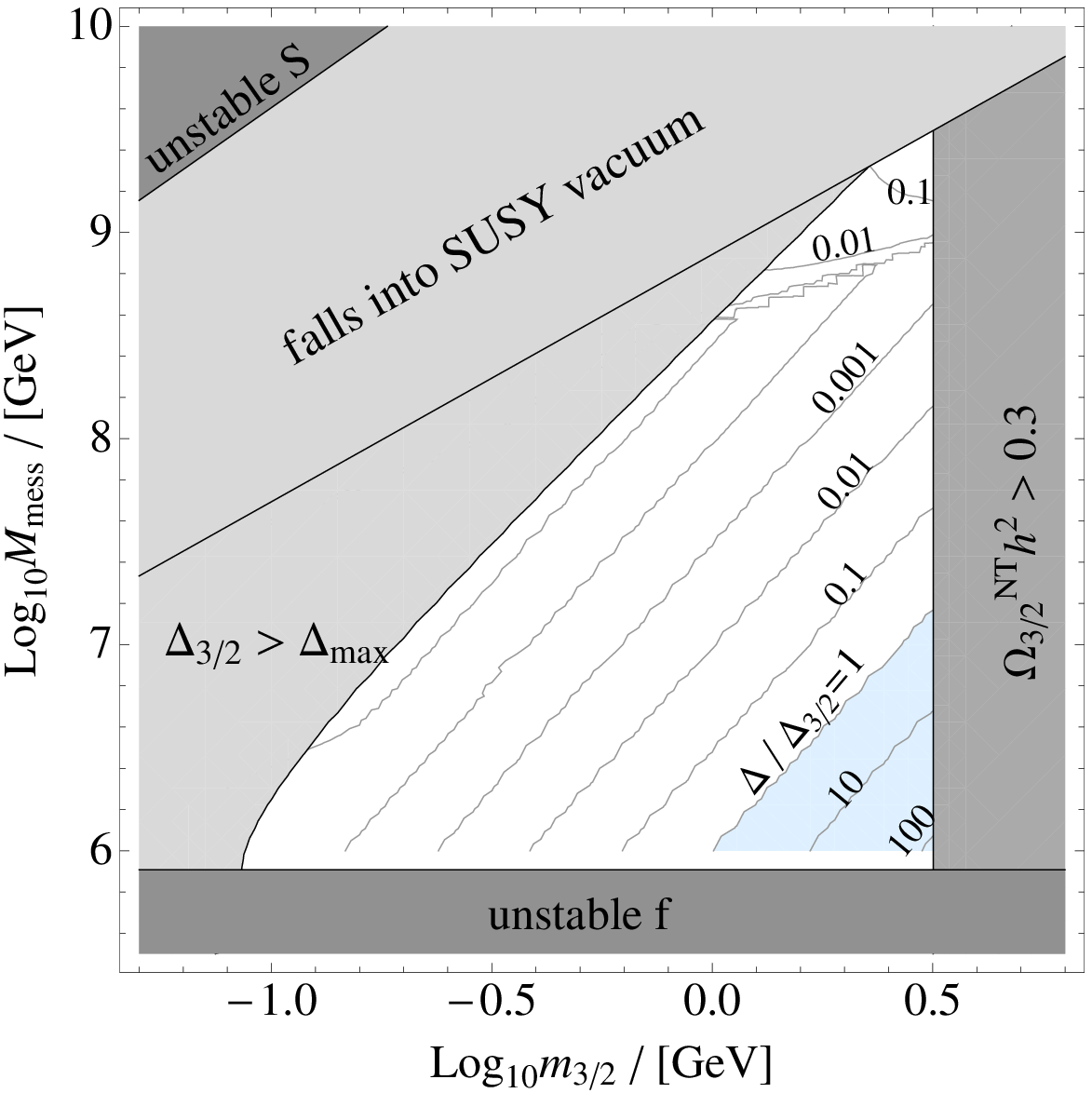}
\includegraphics[width=7cm]{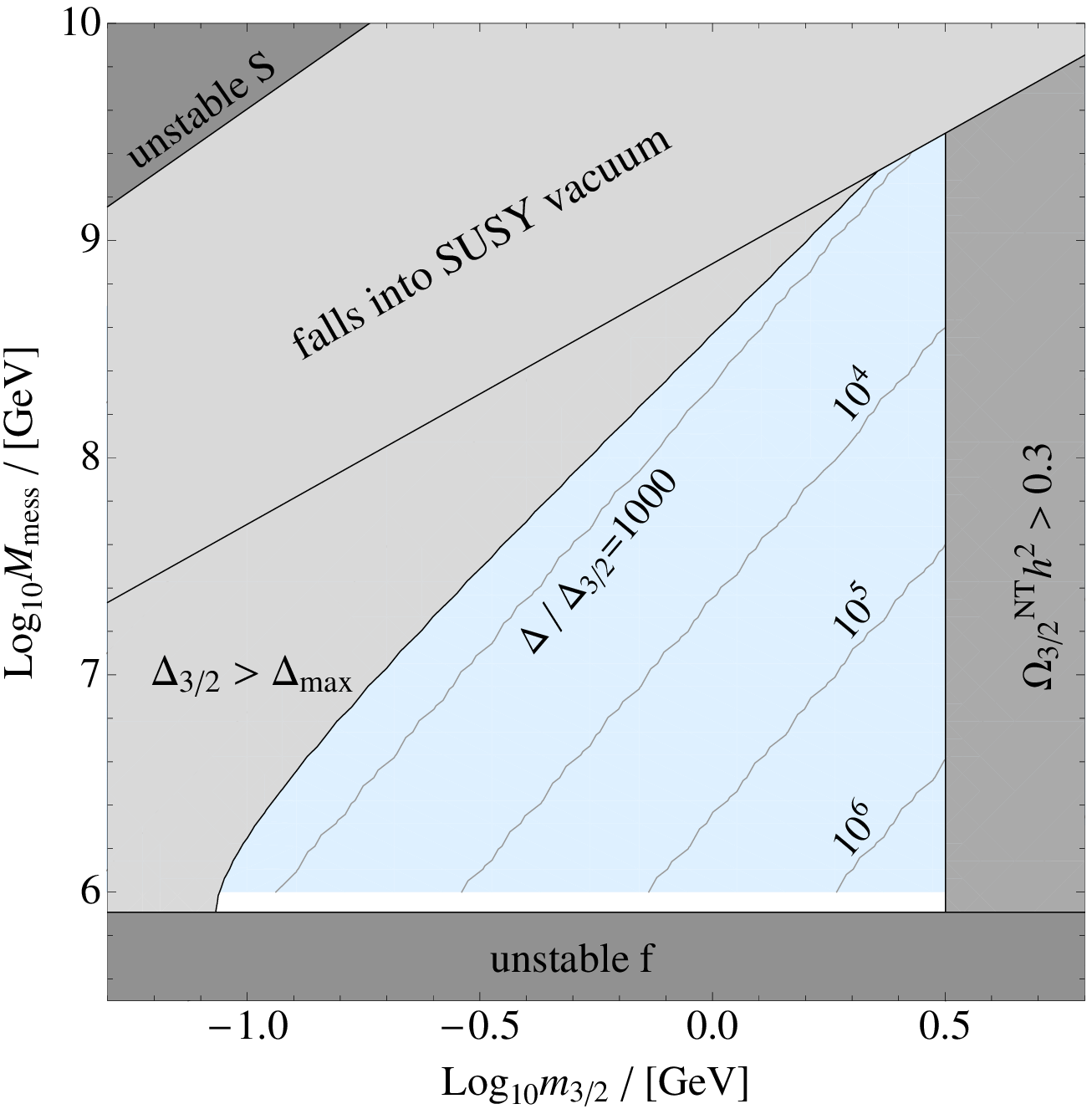}
\caption{The results of numerical study.
In the left(right) figure the initial condition of the position of $S$ right after the inflation 
is taken to be $S_0 = \Lambda$($S_0=M_{\rm pl}$).
The required amount of dilution factor and the theoretical prediction are denoted as $\Delta_{3/2}$ and $\Delta$.
In the blue regions a sizable amount of entropy enough to dilute overabundant gravitino is produced by 
the decay of $S$.
We see that a required amount of dilution factor read off from Eq.~(\ref{delta32}) can be always 
supplied by the decay by choosing an appropriate value of $S_0$ from
between $\Lambda$ and $M_{\rm pl}$.
		 }
\label{dilution_result}
\end{center}
\end{figure}

\subsection{Comments on a light higgsino}

So far we have studied a new cosmological scenario with a high SUSY
scale $M_{\rm SUSY} \gtrsim 5~{\rm TeV}$ in order to realize a 125 GeV
Higgs boson mass. If all the SUSY particles are as heavy as 5~TeV, it is
difficult to confirm the scenario by the LHC experiments.
However, it is possible that the $\mu$-parameter in the MSSM is much
smaller than other superparticle masses. In the GMSB model we used for
the cosmological study there is a natural solution to the $\mu$-problem
(we mention the prescription in appendix~\ref{app:higgsino}).
The model predicts a light higgsino with its mass of
$\mathcal{O}(100)~{\rm GeV}$.  The $\mu$-term is generated by a direct
coupling between SUSY breaking chiral multiplet and Higgs multiplets
assumed at the cutoff scale $\Lambda$, which results in a relatively
small $\mu$-term compared to Higgs soft mass parameters.  
For a cosmologically favorable region of the gravitino mass, the
lightest higgsino does not decay inside the detector. In that case,
searches for mono-jet processes at LHC or mono-photon ones at the ILC
will be able to find the light higgsino.


One should check if a light higgsino scenario is compatible with the
constraint from the BBN. If the higgsino mass is so small that the
life-time becomes as long as $\mathcal{O} (1)$~sec, the decay may alter the
abundance of the light elements.  We have checked the BBN constraints in
the case of $m_{\tilde h} = 300~{\rm GeV}$ and found that such a light
higgsino is cosmologically safe if the gravitino is lighter than $\sim
500$~MeV.  A detailed discussion is given in
appendix~\ref{app:higgsino}.

\section{Summary}

We re-investigated the thermal production of the gravitino in general
framework of gauge mediation.  Calculating the gravitino production
cross section using both the goldstino Lagrangian and the supergravity
one, we confirmed that the relic abundance become insensitive to the
reheating temperature if the temperature of the Universe once exceeds
the messenger mass scale.  Inspired by this property, we presented a new
cosmological scenario; the gravitino dark matter and the thermal
leptogenesis are compatible, namely the ratio $\Omega_{3/2} / \Omega_B$
coincides the observation, $\Omega_{\rm DM} / \Omega_B \sim5$, with an
appropriate value of reheating temperature.  To realize the correct
absolute value of each quantity, $\Omega_{\rm DM} h^2 \simeq 0.1$ and
$\Omega_B \simeq 0.045$, a late-time entropy release is required, which
is automatically supplied by the oscillation energy of the
pseudo-moduli.

To make sure that the scenario actually works, we examined cosmological
evolution of the pseudo-moduli field in a concrete model of gauge
mediation.  With an appropriate initial condition, we showed that the
oscillation energy of the pseudo-moduli dominates the energy density of
the Universe and a sizable amount of entropy needed to fix the energy
densities of gravitino and baryon is released by the subsequent decay.
The scenario is realized when the gravitino mass is $100~{\rm MeV}
\lesssim m_{3/2} \lesssim 1~{\rm GeV}$ and the messenger scale is
$10^6~{\rm GeV} \lesssim M_{\rm mess} \lesssim 10^9~{\rm GeV}$.

Although we have studied the scenario with $M_{\rm SUSY} \gtrsim 5~{\rm
TeV}$ to account for the 125 GeV Higgs boson, the higgsino can be as
light as $\mathcal{O}(100)~{\rm GeV}$.  Such a light higgsino can be
discovered in a future experiments.  We have checked that a light
higgsino is safe from the BBN constraints if the gravitino mass is
smaller than $\sim$~500 MeV.

\section*{Acknowledgments}
We would like to thank Fuminobu Takahashi for discussions on gravitino thermal production.
RK is supported in part by the Grant-in-Aid for Scientific Research
23740165 of JSPS and 25105011 of MEXT.


\appendix

\section{Pseudo-moduli interactions with the MSSM fields}
\label{app:int}

We summarize the interactions between the pseudo-moduli and the MSSM fields 
needed to study the decay of $S$.
The pseudo-moduli interacts with the MSSM fields through the messenger loop diagrams.
The interactions can be read off from the $\langle S \rangle$ dependence of the
 low energy parameters~\cite{Ibe:2006rc,Hamaguchi:2009hy}.
For scalar fields $\tilde{f}$, the effective interaction Lagrangian is written as
\begin{align}
	\mathcal{L}_{\tilde f}
		=
		\frac{(m_{\rm eff}^{\tilde{f}})^2}{\langle S \rangle}
		S \tilde{f}^\dagger \tilde{f} 
		+ {\rm h.c.}
\end{align}
The effective mass parameter $(m_{\rm eff}^{\tilde f})^2$ is a part of the scalar mass that is proportional to
$1/ | \langle S \rangle |^2$.
One element of the scalar mass is the contribution from the gauge mediation,
\begin{align}
	(m_{\rm GM}^{\tilde f})^2 = 
		\left[
			\frac{g^2}{(4 \pi)^2}
		\right]^2
		\cdot
		2 C_2 
		\left|
		\frac{m^2}{\langle S \rangle}
		\right|^2,
	\label{GMscalar}
\end{align}
which is induced at the messenger mass scale $M_{\rm mess}$.
If the gauge mediation is the only source of the scalar mass, $m_{\rm eff}^{\tilde f}$ is identical to their mass.
In that case, $m_{\rm eff}^{\tilde f}$ is the gauge mediation contribution plus the radiative corrections.
In appendix~\ref{app:higgsino}, we consider a direct coupling between the $S$ field and 
the Higgs superfields to solve the $\mu$-problem.
In that case, $m_{H_u}$ consists of two sources;
one is from the gauge mediation and the other is from the direct coupling.
The latter piece does not depend on $\langle S \rangle$,
and has little effect on the effective coupling constant.

As we evaluate the abundance of non-thermally produced higgsino to check the BBN constraint 
in appendix~\ref{app:higgsino},
we list the interaction with higgsino,
\begin{align}
	\mathcal{L}_{\tilde h}
	=
	- \frac{\mu_{\rm eff}}{\langle S \rangle}
	S ~( \bar{h_d^c} \cdot P_L h_u) + {\rm h.c.}
\end{align}
The coefficient $\mu_{\rm eff}$ is again a part of $\mu$ that is proportional to $1/\langle S \rangle$.
Actually, as we see in appendix~\ref{app:higgsino},
$\mu$-term is generated at the cutoff scale $\Lambda$ through 
the K$\ddot{\rm a}$hler potential Eq.~(\ref{Ksweet}) and it does not have $\langle S \rangle$ dependence.
The VEV dependence of $\mu_{\rm eff}$ appears only through the renormalization group running,
and the effect is very small for the $\mu$-term.
The effective coupling $\mu_{\rm eff}$ is suppressed compared to  the $\mu$-term, typically
\begin{align}
	| \mu_{\rm eff} |
	\sim
	0.01 \times
	| \mu |.
\end{align}

Among the effective couplings, the Higgs mass parameter $m_{\rm eff}^{H_u}$ is 
enhanced by the large renormalization group running~\cite{Hamaguchi:2009hy},
\begin{align}
	- (m_{\rm eff}^{H_u} )^2 = ( \kappa m_{\tilde B})^2,
\end{align}
with
\begin{align}
	\kappa \simeq 3-4.
\end{align}

\noindent
{\bf Decays of the $S$ field}

The $S$ field mainly decays into the MSSM particles.
Since the mass parameter $m_{\rm eff}^{H_u}$ is enhanced over other SUSY breaking parameters,
the decay rate into the Higgs boson is enhanced.
For $m_S > 2 m_h$, the main decay channel turns out to be
$S \to hh, ZZ$ and $WW$, where the gauge bosons are longitudinally polarized~\cite{Hamaguchi:2009hy},
\begin{align}
	\Gamma_{S \to hh} + \Gamma_{S \to ZZ} + \Gamma_{S \to WW}
		\simeq
		\frac{1}{8 \pi m_S}
		\left(
			\frac{(m_{\rm eff}^{H_u})^2 \sin^2 \beta}{\langle S \rangle}
		\right)^2.
	\label{Gammamain}
\end{align}
Approximating the total decay width $\Gamma_S$ by that of main channel,
the decay temperature defined in Eq.~(\ref{Td}) is written as 
\begin{align}
	T_d
	\simeq
	68 {\rm GeV}
	\left(
		\frac{g_\ast}{15}
	\right)^{-1/4}
	\Bigl(
		\frac{m_{\rm eff}^{H_u}}{5 ~{\rm TeV}}
	\Bigr)^2
	\left(
		\frac{m_{\tilde g}}{5 ~{\rm TeV}}
	\right)^{3/4}
	\left(
		\frac{m_{3/2}}{500 ~{\rm MeV}}
	\right)^{-5/4}.
	\label{Tdmain}
\end{align}

There is also a rare decay mode $S \to \psi_{3/2} \psi_{3/2}$, which become important if the gravitino mass 
is larger than $\sim 1~{\rm GeV}$.
The decay width is calculated to be~\cite{Ibe:2006rc,Hamaguchi:2009hy}
\begin{align}
	\Gamma_{3/2}
	=
	\frac{1}{96 \pi} \frac{m_S^3}{M_{\rm pl}^2}
	\left(
		\frac{m_S}{m_{3/2}}
	\right)^2.
\end{align}
If $S$ dominates the energy density of the Universe, non-thermal gravitino abundance is calculated to be
\begin{align}
	\Omega_{3/2}^{\rm NT}
	=
	\frac{3}{4} m_{3/2}
	\frac{T_d}{m_S}
	\times 2 B_{3/2}
	/
	(\rho_c / s)_0,
\end{align}
where $(\rho_c / s)_0 \simeq 1.8 \times 10^{-9} {\rm GeV}$ is the critical  density divided by the entropy density
at present.
Approximating the decay temperature as Eq.~(\ref{Tdmain}), the non-thermal gravitino is estimated as
\begin{align}
	\Omega_{3/2}^{\rm NT} \simeq
	0.2 
	\left(
		\frac{m_{3/2}}{2~{\rm GeV}}
	\right)^{9/4}
	\left(
		\frac{m_{\tilde g}}{5~{\rm TeV}}
	\right)^{5/4}
	\Bigl(
		\frac{m_{\rm eff}^{H_u}}{5~{\rm TeV}}
	\Bigr)^{-2}.
\end{align}
We have used the formula in Fig.~\ref{delta_TR}.

\section{$\mu$-problem and a light higgsino}
\label{app:higgsino}

Here we present a possible solution to the $\mu$-problem.
As we see below, the solution predicts a relatively light higgsino compared to $M_{\rm SUSY}$.
We check whether a light higgsino scenario is allowed by the BBN constraint.

In order to avoid too large $\mu$-term,
we assume an approximate Peccei-Quinn (PQ) U(1) symmetry with a charge assignment $PQ(H_u)=PQ(H_d)=1$.
Also, to realize the relation $\mu^2 \sim m_{H_u}^2$, 
we assume the following general interactions between $S$ and the Higgs superfields at the cutoff scale~\cite{Ibe:2007km},
\begin{align}
	K^{(\rm{Higgs})}
		=
		\left(
			c_\mu \frac{S^\dagger H_u H_d}{\Lambda} + \rm{h.c.}
		\right)
		-
		c_H \frac{S^\dagger S (H_u^\dagger H_u + H_d^\dagger H_d)}{\Lambda^2},
	\label{Ksweet}
\end{align}
where the PQ charge of $S$ is fixed as $PQ(S)=2$.
Once the $F$-component of $S$ develops a VEV, 
$\mu$-term and the Higgs scalar mass terms emerge at the scale $\Lambda$.
The relation $\mu^2 \sim m_{H_u}^2$,
which is needed for satisfying the condition of electroweak symmetry breaking without a serious fine-tuning, 
naturally realizes if the coefficients $c_\mu$ and $c_H$ are both $\mathcal{O} (1)$.

Possible origins of the K$\ddot{\rm a}$hler potential~(\ref{Ksweet}) are discussed in Ref.~\cite{Ibe:2007km}
by studying dynamics of UV models above the cutoff scale $\Lambda$.
There, it is found that the coefficients $c_\mu$ and $c_H$ tend to have a mild hierarchy,
and we typically have $\mu / m_H \sim 1/10$.
This hierarchy implies that the Higgs scalar mass parameter $m_{H_u}$ tends to be above the order of TeV scale 
for a moderate value of $\mu$-term,
namely $m_{H_u} \gtrsim \mathcal{O} (1)~{\rm TeV}$ for 
$\mu \gtrsim \mathcal{O} (100)~{\rm GeV}$.

We do not regard this small hierarchy as catastrophic; actually, 
this hierarchy is consistent with the relatively heavy Higgs boson mass.
In order for the electroweak symmetry to be broken radiatively, the condition
\begin{align}
	\frac{M_Z^2}{2}
	\simeq
	-\mu^2 - m_{H_u}^2 (\Lambda) - \delta m_{H_u}^2
	\label{ewbcondition}
\end{align}
must be satisfied. $\delta m_{H_u}^2$ is a contribution from the radiative corrections.
With positive $m_{H_u}^2 (\Lambda)$ and $\mu^2 \ll m_{H_u}^2 (\Lambda)$,
$\delta m_{H_u}^2$ must be negative and large to  satisfy the condition~(\ref{ewbcondition}),
which is realized by the contributions from the stop-loop diagrams if the stop mass $m_{\tilde t}$ is large.
Large stop mass subsequently induce a large contribution proportional to $m_{\tilde t}^2$
to the Higgs boson mass again through the stop-loop diagram to realize a relatively heavy 
Higgs boson.
In summary, in this set-up, the $\mu$-problem is ameliorated by the generalized version of the Giudice-Masiero mechanism 
with the K$\ddot{\rm a}$hler potential in Eq.~(\ref{Ksweet}), which in turn leads the relatively small $\mu$-term and the relatively heavy Higgs boson mass in accord with $m_h = 125 ~{\rm GeV}$.

Although it is difficult to discover a SUSY particles at the LHC experiments
when $M_{\rm SUSY} \sim 5 ~{\rm TeV}$,
it predicts a light higgsino with $m_{\tilde h} \simeq \mathcal{O} (100) {\rm GeV}$.
Therefore, in this scenario, there is a chance 
to discover a light higgsino in the future experiment.


The light higgsino in GMSB is subject to the constraints from BBN.
The constraints on the primordial abundance of the lightest neutralino $\chi$ is studied in Ref.~\cite{Kawasaki:2008qe}.
They analyzed the decay process of the neutralino and presented constraints on $Y_\chi = n_\chi / s$, 
the yield of $\chi$, in a Bino-like NLSP case.
We use the constraints to derive those for the higgsino.

Since the life-time of a neutralino $\chi$ is approximately proportional to
$m_{3/2}^2 /  m_{\chi}^5$,
constraints on the primordial  abundance are more severe 
for larger $m_{3/2}$ or smaller $m_{\chi}$.
We focus on a case that the mass of NLSP (in our case higgsino) is $300 ~{\rm GeV}$.
According to Ref.~\cite{Kawasaki:2008qe},
if the gravitino is heavier than $\sim 500 ~{\rm MeV}$, the stringent bound on the bino abundance comes from
the overproduction of the Deuterium.
For $10 ~{\rm MeV} \lesssim m_{3/2} \lesssim 500 ~{\rm MeV}$,
the bound is from the overproduction of $^4 {\rm He}$,
\begin{align}
	&m_{\tilde B} Y_{\tilde B} \lesssim 10^{-13} {\rm GeV} \ \ \
		(500 ~{\rm MeV} \lesssim m_{3/2} \lesssim 100~{\rm GeV}), \label{bbn1}
	\\
	&m_{\tilde B} Y_{\tilde B} \lesssim 10^{-9} {\rm GeV} \ \ \ \
		(10 ~{\rm MeV} \lesssim m_{3/2} \lesssim 500 ~{\rm MeV}). \label{bbn2}
\end{align}
The bound is much weaker for $m_{3/2} \lesssim 10 ~{\rm MeV}$.
We estimate the higgsino abundance in the scenario and check whether a light higgsino is allowed 
by BBN.

Higgsinos are produced non-thermally from the decay of the pseudo-moduli,
\begin{align}
	Y_{\tilde h} =
	\frac{3}{4} \frac{T_d}{m_S}
	\times 2 B_{\tilde h},
\end{align}
where $B_{\tilde h}$ is the branching ratio of the decay process $S \to \tilde{h} \tilde{h}$
and the decay temperature $T_d$ is well approximated by Eq.~(\ref{Tdmain}).
$Y_{\tilde h}$ depends on two effective couplings :
$m_{\rm eff}^{H_u}$ and $\mu_{\rm eff}$ defined in appendix~\ref{app:int}.
Remaining these parameters, the higgsino abundance is estimated as 
\begin{align}
	m_{\tilde h} Y_{\tilde h}
	\simeq
	1.2 \times 10^{-7} {\rm GeV}
	\left(
		\frac{m_{3/2}}{500~{\rm MeV}}
	\right)^{-3/4}
	\Bigl(
		\frac{m_{\rm eff}^{H_u}}{5~{\rm TeV}}
	\Bigr)^{-2}
	\left(
		\frac{\mu_{\rm eff}}{5~{\rm GeV}}
	\right)^2.
\end{align}

The abundance of the non-thermally produced higgsinos is decreased by the subsequent annihilation process.
This effect can be taken into account by solving the Boltzmann equation,
\begin{align}
	\dot{n}_{\tilde h} + 3 H n_{\tilde h} =
	- \langle \sigma v\rangle n_{\tilde h}^2,
	\label{Boltzhiggsino}
\end{align}
where $\langle \sigma v \rangle$ is 
the thermal averaged annihilation cross section of higgsino~\cite{ArkaniHamed:2006mb}
\footnote[2]{We have not included co-annihilation effects to make a conservative estimate.}
,
\begin{align}
	\langle \sigma v \rangle 
	=
	\frac{g^4}{128 \pi \mu^2} 
	\left(
		\frac{3}{2} + \tan^2 \theta_W + \frac{\tan^2 \theta_W}{2}
	\right),
\end{align}
where $\theta_W$ is Weinberg angle.
The solution of the Boltzmann equation~(\ref{Boltzhiggsino}) is approximated by a simple analytic
formula~\cite{Fujii:2002kr,Hamaguchi:2009hy}.
In terms of the yield value $Y_{\tilde h} = n_{\tilde h} /s$,
\begin{align}
	Y_{\tilde h} (T)
	\simeq
	\left[
		\frac{1}{Y_{\tilde h} (T_d)}
		+
		\ro{\frac{8 \pi^2 g_\ast (T_d)}{45}}
		\langle \sigma v \rangle M_{\rm pl} (T_d -T)
	\right]^{-1}.
\end{align}
If the initial abundance $Y_{\tilde h} (T_d)$ produced by the decay of $S$ is large enough,
the resultant abundance for $T \ll T_d$ is independent of $Y_{\tilde h} (T_d)$.
In this case, the abundance is estimated by
\begin{align}
	Y_{\tilde h} \simeq
	8.2 \times 10^{-13}
	\left(
		\frac{15}{g_\ast}
	\right)^{1/2}
	\left(
	\frac{10 ~{\rm GeV}}{T_d}
	\right)
	\left(
	\frac{10^{-8} ~{\rm GeV}^{-2}}{\langle \sigma v \rangle}
	\right).
\end{align}
For higgsino with $m_{\tilde h} = 300~{\rm GeV}$, 
\begin{align}
	m_{\tilde h} Y_{\tilde h}
	\simeq
	3.9 \times 10^{-11} {\rm GeV}
	\Bigl(
		\frac{15}{g_\ast}
	\Bigr)^{3/4}
	\Bigl(
		\frac{m_{\rm eff}^{H_u}}{5~{\rm TeV}}
	\Bigr)^{-2}
	\left(
		\frac{m_{\tilde g}}{5~{\rm TeV}}
	\right)^{-3/4}
	\left(
		\frac{m_{3/2}}{500~{\rm MeV}}
	\right)^{5/4}
	\left(
		\frac{\mu}{300~{\rm GeV}}
	\right)^2.
\end{align}
Compared with Eq.~(\ref{bbn1}) and (\ref{bbn2}), we see that 
the higgsino abundance is below the BBN constraint for $m_{3/2} \lesssim 500~{\rm MeV}$
with the help of the annihilation process.




\end{document}